\documentclass[paper,12pt]{JHEP3}
\usepackage{epsfig}
\usepackage{amsmath}
\usepackage{amssymb}

\title{Tomography of cold and hot QCD matter : \\[0.5em] tools and diagnosis}

\author{Fran\c{c}ois Arleo\\ ECT* and INFN, G.C. di Trento,\\ Strada delle Tabarelle, 286\\38050 Villazzano (Trento), Italy \\E-mail: \email{francois@ect.it}}

\abstract{The probability distribution $D(\eps)$ in the energy loss incurred by incoming and outgoing hard quarks in a QCD medium is computed numerically from the BDMPS gluon spectrum. It is shown to follow an empirical log-normal behavior which allows us to give the quenching weight a simple analytic parameterization. The dependence of our results under the infrared and ultraviolet sensitivity of the gluon spectrum is investigated as well. Finally, as an illustration, we discuss and compare estimates for the quenching of hadron spectra in nuclear matter and in a quark-gluon plasma to HERA and RHIC preliminary data.}

\keywords{QCD, Jets, Hadronic Colliders}

\preprint{JHEP {\bf 11} (2002) 044}

\def\abs#1{\left|#1\right|}
\def\cO#1{{{\cal{O}}}\left(#1\right)}

\newcommand\eps{\epsilon}
\newcommand\Dbar{\bar{D}}
\newcommand\Dt{\tilde{D}}
\newcommand\Et{{\bar{E}}}
\newcommand\epsb{{\bar{\epsilon}}}
\newcommand\ombar{{\bar{\omega}}}

\begin{document}

%%%%%%%%%%%%%%%%%%%%%%%%%%%%%%%%%%%%%%%%%%%%%%%%%%%%%%%%%%%%%%%%%%%%
\section{Introduction}
%%%%%%%%%%%%%%%%%%%%%%%%%%%%%%%%%%%%%%%%%%%%%%%%%%%%%%%%%%%%%%%%%%%%

The energy loss experienced by a fast parton may serve as a measure of the density of color charges of the QCD medium it travels through~\cite{G:1990yeG:1992xbG:1994hr,Baier:1995bd}. It may be huge in a dense medium such as the quark-gluon plasma (QGP), expected to be formed in the early stage of ultrarelativistic heavy ion collisions. One of the most direct and observable consequences would be the quenching of high $p_\perp$ particle production~\cite{Wang:1992xy,Baier:1997kr}. This is reported experimentally. The first PHENIX data showed a large suppression in central Au-Au collisions as compared to proton-proton measurements~\cite{Adcox:2001jp}, later confirmed on a wider $p_\perp$ range by the PHENIX~\cite{Adcox:2002pe}, PHOBOS~\cite{PHOBOS}, and STAR~\cite{Adler:2002xw} collaboration. The quenching factor of $Q(p_\perp) \approx 5$ observed at large $p_\perp$ finds therefore a natural explanation in this context. This makes possible tomographic investigations~\cite{Gyulassy:2001zv}, i.e., the study of the (color) structure of hot QCD matter by measuring the energy attenuation of hard probes.

Although the understanding of medium-induced parton energy loss has been extensively developed over the last few years~\cite{G:1990yeG:1992xbG:1994hr,Baier:1995bd,Wang:1995fx,Baier:1997sk,Baier:1998kq,Z:1996fvZ:1997uu,G:1999zdG:2000fsG:2000er,W:2000tfW:2000za}, less is known about how to relate this mechanism to observable quantities. A step in that direction has however been taken recently by Baier, Dokshitzer, Mueller, and Schiff (BDMS) in Ref.~\cite{Baier:2001yt} in which they connect the measured quenching factor $Q(p_\perp)$ to the induced gluon spectrum $dI/d\omega$ radiated by the leading parton. Perhaps even more importantly, the authors emphasize that the standard modeling of the quenching --~determined by the {\it mean} energy loss~-- proves inadequate. Rather, the knowledge of the full probability distribution $D(\eps)$ in the energy loss is actually required. On general grounds, the cross sections will be modified in the medium as
\begin{equation}\label{eq:mediumxs}
\sigma^{medium} = D(\eps)\,\otimes\,\sigma^{vacuum}
\end{equation}
However, unless strong assumptions are made as for $dI/d\omega$,  $D(\eps)$ cannot be calculated analytically in a simple way. It is therefore the aim of this paper to present and discuss a numerical computation of the distribution $D(\eps)$ from the medium-induced gluon spectrum derived by Baier, Dokshitzer, Mueller, Peign\'e, and Schiff (BDMPS). Using these results, we give some estimates for the quenching (``diagnosis'') to be compared to experimental data in electron-nucleus and nucleus-nucleus collisions.

The outline of the paper is as follows. In Section~\ref{se:setup}, we detail the numerical procedure to compute $D(\eps)$ from its integral representation given in~\cite{Baier:2001yt}. The distribution is simply related to the gluon multiplicity $N(\omega)$ radiated by the leading parton, calculated in Sect.~\ref{subse:mult}. Results are given in Section~\ref{se:results}. The probability distributions for both an incoming and an outgoing quark are first presented (Sect.~\ref{subse:inout}) and then contrasted with analytical approximations (Sect.~\ref{subse:comp}). A simple analytical parameterization of the quenching weight $D(\eps)$ ends the Section. We examine in Section~\ref{se:approx} the sensitivity of our results beyond the approximations made in Section~\ref{se:results} for the infrared and ultraviolet behavior of the BDMPS medium-induced gluon spectrum. We apply our results in Section~\ref{se:app} where the quenching of hadron spectra is determined in cold and hot QCD matter and compared qualitatively to HERMES and PHENIX preliminary data. Finally, Section~\ref{se:summary} summarizes the main results of the present study.

%%%%%%%%%%%%%%%%%%%%%%%%%%%%%%%%%%%%%%%%%%%%%%%%%%%%%%%%%%%%%%%%%%%%
\section{Computation of the energy loss distribution $D(\epsilon)$}\label{se:setup}
%%%%%%%%%%%%%%%%%%%%%%%%%%%%%%%%%%%%%%%%%%%%%%%%%%%%%%%%%%%%%%%%%%%%

\subsection{Integral representation} 

The multiple soft collisions undergone by a hard parton traveling through a medium induce gluon emission. Consequently, these radiated gluons take away an energy $\epsilon$ from the leading particle with a probability distribution (or quenching weight) $D(\eps)$. Let us briefly recall here its expression found by BDMS~\cite{Baier:2001yt}.

As long as interference effects between radiated gluons (suppressed by $\alpha_S$) can be neglected, we may assume that the gluon emissions from the leading parton are independent\footnote{Strictly speaking, one has to assume moreover that the energy loss $\eps$ remains much smaller than the parton energy $\eps \ll E$ (soft limit), which is precisely the case of interest in this section. We shall come back to this point in Section~\ref{se:approx} where a finite quark energy will be explicitly considered.}. This allows for a Poisson formulation of $D(\eps)$, which reads~\cite{Baier:2001yt}
\begin{equation}  \label{eq:poisson}
D(\eps) = \sum^\infty_{n=0} \, \frac{1}{n!} \,
\left[ \prod^n_{i=1} \, \int \, d\omega_i \, \frac{dI(\omega_i)}{d\omega} 
\right] \delta \left(\eps - \sum_{i=1}^n  \omega_i\right)
\,\cdot \,\exp \left[ - \int_0^{+\infty} d\omega \, \frac{dI(\omega)}{d\omega} \right]. 
\end{equation} 
Here, $dI / d\omega$ represents the medium-induced gluon spectrum and $n$ the number of radiated gluons by the hard parton. Note that secondary gluon emissions, neglected in the soft limit, are not taken into account in~(\ref{eq:poisson}).

Using the Mellin representation of the delta function, the series~(\ref{eq:poisson}) is resumed to finally obtain an integral representation
\begin{equation}\label{eq:laplace}
D(\eps) = \int_{\cal{C}} \, \frac{d\nu}{2\pi i} \>
\tilde{D}(\nu)\,e^{\nu {\eps}}, 
\end{equation}
with the integration contour $\cal{C}$ chosen (here) to be the imaginary axis. The Laplace transform $\Dt(\nu)$ is simply related to the induced gluon spectrum through
\begin{equation}
\label{eq:dtilde}
 \tilde{D}(\nu)  = \exp \left[ -\nu\int^\infty_0 d\omega\,e^{-\nu\omega} \, 
N\left(\omega\right) \right] 
\end{equation}
where the integrated gluon multiplicity $N(\omega)$ is defined as the number of gluons with an energy larger than $\omega$, i.e.,
\begin{equation}\label{eq:mult}
  N\left( {\omega}\right) \equiv \int_\omega^\infty d\omega'\, \frac{dI(\omega')}{d\omega'}.
\end{equation}
Taking $\nu= i\,b$ in (\ref{eq:laplace}), the distribution $D(\eps)$ thus becomes
\begin{equation}\label{eq:laplace2}
D(\epsilon)=\int_0^{+\infty}\,\frac{db}{\pi} \,\exp\left(-b\,I_s(b)\right)\,\cos\left(b(\epsilon-I_c(b))\right)
\end{equation}
with
\begin{equation}\label{eq:I}
\begin{split}
I_c(b) & = \int_0^{+\infty}\,d\omega\,\cos\left(b\,\omega\right)\,N(\omega), \\
I_s(b) & = \int_0^{+\infty}\,d\omega\,\sin\left(b\,\omega\right)\,N(\omega). 
\end{split}
\end{equation}
Given a medium-induced gluon spectrum $dI/d\omega$, the integrated gluon multiplicity $N(\omega)$ can be determined exactly. This then allows for the numerical computation of the integrals (\ref{eq:I}) and, subsequently, of the quenching weight $D(\eps)$ through Eq.~(\ref{eq:laplace2}). The multiplicity of soft radiated gluons computed from the BDMPS spectrum for outgoing and incoming quarks is now discussed.

\subsection{Gluon multiplicities}\label{subse:mult}

\vspace{0.4cm}
{\large {\it Outgoing quarks}}
\vspace{0.4cm}

Let us start with the BDMPS gluon spectrum radiated by an outgoing quark with energy $E$ traversing a medium of length $L$. In the soft gluon approximation ($\omega \ll E$), it reads~\cite{Baier:1997kr,Baier:1998kq}
\begin{eqnarray}
\label{eq:dIdo_out}
\frac{dI(\omega)}{d\omega} &=&\frac{\alpha}{\omega} \ln
 \abs{\cos\left[\,(1+i) u \,\right]} \\
&=& \frac{\alpha}{2\omega} \,
\ln \left[\,\cosh^2 u - \sin^2 u \,\right] ; 
\qquad u \equiv \sqrt{\frac{\omega_c}{2\,\omega}} \quad,\quad
\alpha\equiv \frac{2\alpha_s C_R}{\pi}. \nonumber 
\end{eqnarray}
where $C_F = 4/3$ is the Casimir operator in the fundamental representation and $\alpha_s = g^2 / 4\pi \simeq 1/2$ the strong coupling constant . In the soft limit, the spectrum~({\ref{eq:dIdo_out}) is thus characterized by only one energy scale
\begin{equation}\label{eq:omc}
\omega_c = \frac{1}{2}\,\hat{q}\,L^2
\end{equation}
where the so-called gluon transport coefficient $\hat{q}$ measures the ``scattering power'' of the medium. Relating it to the gluon density of the medium, BDMPS give perturbative estimates for the transport coefficient. While it is shown to be as small as $\hat{q} \simeq 0.25$~GeV/fm$^2$ in cold nuclear matter, a much larger $\hat{q} \simeq 5$~GeV/fm$^2$ is expected in a hot ($T = 250$~MeV) quark-gluon plasma~\cite{Baier:1997sk}.

Writing $N(\omega)$ as 
\begin{equation}\label{eq:multan}
  N\left( {\omega}\right) = \int_0^\infty d\omega'\, \frac{dI(\omega')}{d\omega'} - \int_0^\omega d\omega'\, \frac{dI(\omega')}{d\omega'}
\end{equation}
and Taylor expanding the gluon spectrum at\footnote{We shall keep this notation for all variables normalized to the energy scale $\omega_c$.} $\ombar \equiv \omega/\omega_c \ll 1$, the gluon multiplicity reads~\cite{Baier:2001yt}
\begin{equation}\label{eq:Nsmall_out}
 N(\omega \ll \omega_c) \simeq \alpha\left(\,  \sqrt\frac{2}{\ombar} +\ln2\ln \ombar - 1.44136 \right)
\end{equation}
in the infrared (IR) domain, while it drops much faster 
\begin{equation}\label{eq:Nlarge_out}
N(\omega \gg \omega_c) = \frac{\alpha}{24}\, \left(\frac{1}{\ombar}\right)^2
\end{equation}
in the ultraviolet (UV) region. It is however necessary to rely on a numerical determination of $N(\omega)$ in between these two regimes. We decompose therefore the integrals~(\ref{eq:I}) as
\begin{equation}\label{eq:decompose}
I_{c} = \left[ \,\int_0^{\omega_{-}} \qquad + \,\int_{\omega_{+}}^{\infty} \qquad + \,\int_{\omega_{-}}^{\omega_{+}} \, \right] \quad d\omega\,\cos\left(b\,\omega\right)\,N(\omega)
\end{equation}
where the first two terms on the r.h.s. of ({\ref{eq:decompose}) are expressed analytically in terms of Fresnel integrals and hypergeometric functions\footnote{Even though the gluon multiplicity in the BDMPS formalism is sensitive in the IR, all integrals here are infrared safe quantities.} while the third term is determined numerically. The cutoff $\omega_-$ ($\omega_+$) under (above) which the exact gluon multiplicity may be replaced by its analytical approximation (\ref{eq:Nsmall_out}) (respectively, (\ref{eq:Nlarge_out})) is taken to be $\omega_{-} = 0.01\,\omega_c$ ($\omega_{+} = 2\,\omega_c$).

The gluon multiplicity radiated by an outgoing quark $N(\omega)$ is plotted in Figure~\ref{fig:mult} as a function of $\ombar$ ({\it upper solid} line), together with its analytic approximation (\ref{eq:Nsmall_out}) at small energy ({\it dashed}). Anticipating the discussion in Section~\ref{subse:comp}, the approximation 
\begin{equation}\label{eq:Nsmallsmall}
N(\ombar) = \alpha\,  \sqrt\frac{2}{\ombar}
\end{equation}
based on the small $\ombar$ behavior of (\ref{eq:Nsmall_out}) is shown as a dash-dotted line. While the expression ({\ref{eq:Nsmall_out}) is shown to reproduce the exact result for gluon energies up to $\omega \approx 0.3 \,\omega_c$, Eq.~(\ref{eq:Nsmallsmall}) strongly overestimates $N(\omega)$ over the whole range of interest.

\vspace{0.4cm}
\noindent {\large {\it Incoming quarks}}
\vspace{0.4cm}

A similar procedure for the case of an incoming quark can be carried out. The medium-induced gluon spectrum now reads~\cite{Baier:1998kq}
\begin{eqnarray}
\label{eq:dIdo_in}
\frac{dI(\omega)}{d\omega} &=&\frac{\alpha}{\omega} \,\ln
 \abs{\frac{\sin\left[\,(1+i) u\,\right]}{(1+i) u}} \nonumber  \\
&=& \frac{\alpha}{2\omega} \, \ln \left[ \frac{ \cosh^2 u -
\cos^2 u \,}{2 u^2} \right]. 
\end{eqnarray}
As previously, the limiting energy behavior can be extracted analytically. We found
\begin{equation}\label{eq:Nsmall_in}
N(\omega \ll \omega_c ) \simeq \alpha\left(\,  \sqrt\frac{2}{\ombar} + \ln2\ln \ombar - \frac{1}{4}\,\left(\ln \ombar\right)^2 - 1.32099 \right).
\end{equation}
and 
\begin{equation}\label{eq:Nlarge_in}
N(\omega \gg \omega_c) = \frac{\alpha}{360}\, \left(\frac{1}{\ombar}\right)^2
\end{equation}
at small and high energies respectively. The integrals ({\ref{eq:I}) are computed as for the outgoing quark case.

The energy dependence of the gluon multiplicity emitted by the incoming quark is also shown in Figure~\ref{fig:mult} ({\it lower solid} line) together with the analytical expression (\ref{eq:Nsmall_in}) as a dashed line. The number of emitted gluons remains much smaller than what is observed for the outgoing quarks (note the factor 15 difference between (\ref{eq:Nlarge_out}) and (\ref{eq:Nlarge_in})). Indeed, for a hard quark produced {\it in} the medium, a gluon can be emitted shortly after the hard process and prior the first scattering of the hard quark in the medium~\cite{Baier:1998kq,Qiu:2001hj}. This increases the gluon multiplicity (hence, the energy loss) radiated by outgoing quarks produced in the medium. 

\begin{figure}[htbp]
\begin{center}
\includegraphics[width=9.cm]{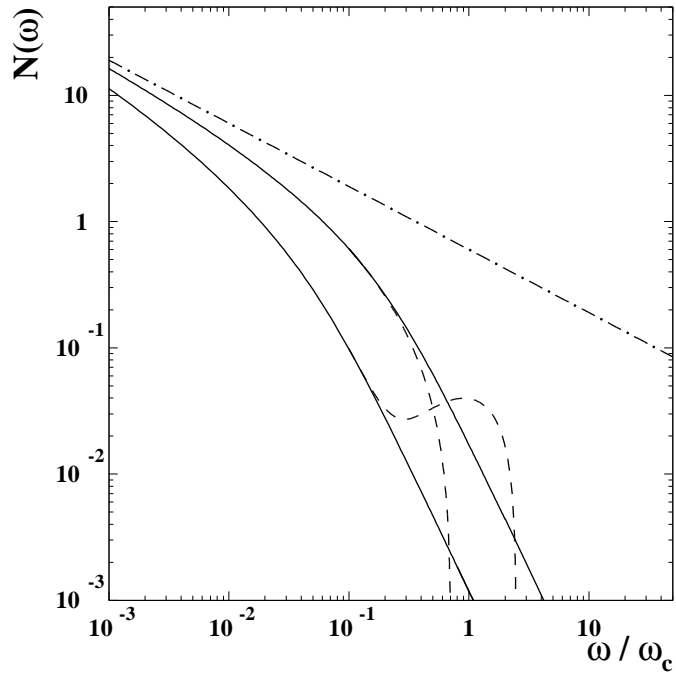}
\caption{Integrated gluon multiplicity $N(\omega)$ computed from the BDMPS spectrum respectively for outgoing ({\it upper solid}) and incoming ({\it lower solid}) quarks. Its analytic IR behavior (resp. Eq.~(\ref{eq:Nsmall_out}) and (\ref{eq:Nsmall_in})) is shown in dashed lines while the dash-dotted line represents the small $\ombar$ approximation~(\ref{eq:Nsmallsmall}), $N(\omega) \propto \ombar^{-1/2}$.}
\label{fig:mult}
\end{center}
\end{figure}

%%%%%%%%%%%%%%%%%%%%%%%%%%%%%%%%%%%%%%%%%%%%%%%%%%%%%%%%%%%%%%%%%%%%
\section{Numerical results}\label{se:results}
%%%%%%%%%%%%%%%%%%%%%%%%%%%%%%%%%%%%%%%%%%%%%%%%%%%%%%%%%%%%%%%%%%%%

\subsection{Incoming vs. outgoing quarks}\label{subse:inout}

Following the procedure detailed in the former Section, the probability $D(\eps)$ that an outgoing or incoming quark loses an energy $\epsilon$ while going through the medium is determined numerically. 

The normalized distribution 
\begin{equation}
\Dbar(\epsb = \epsilon / \omega_c) = \omega_c\,D(\eps)
\end{equation}
is represented in Figure~\ref{fig:dist_inout} for the outgoing ({\it solid line}) and the incoming ({\it dashed}) quark case. It exhibits a strong peak for energy loss $\eps$ much smaller than the typical scale $\omega_c$. Moreover, the long energy tail of the distribution makes the mean energy loss $\langle \epsilon \rangle$ 
\begin{equation}
\langle \epsilon \rangle = \int_0^{+\infty}\,d\eps\,\eps\,D(\eps)
\end{equation}
well larger than its peak value, as emphasized in~\cite{Baier:2001yt}.

\begin{figure}[htbp]
\begin{center}
\includegraphics[width=9.cm]{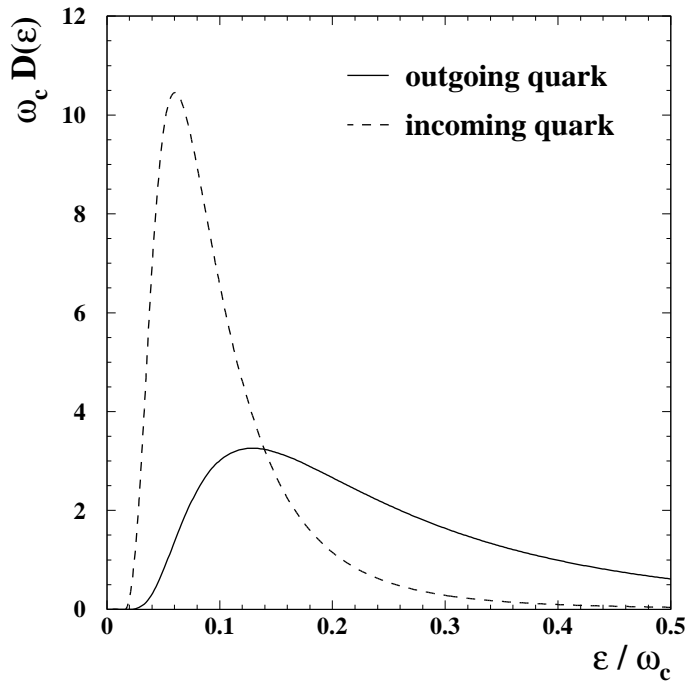}
\caption{Distribution in the energy loss $\Dbar(\epsb)$ for outgoing ({\it solid}) and incoming ({\it dashed}) quarks computed from the BDMPS spectrum in the soft limit.}
\label{fig:dist_inout}
\end{center}
\end{figure}

Furthermore, Figure~\ref{fig:dist_inout} shows that the energy loss suffered by an incoming quark proves much smaller than what happens to a quark produced in the medium. This was already apparent from the gluon multiplicities (Figure~\ref{fig:mult} and end of Sect.~\ref{subse:mult}): the lower the number of emitted gluons for all $\omega$, the smaller the energy loss $\eps$. 

\subsection{Comparison with the BDMS estimate}\label{subse:comp}

The numerical results we obtained previously are compared with analytical estimates for the quenching weight. Let us begin with the illustrative guess based on the small energy behavior of the gluon multiplicity (\ref{eq:Nsmallsmall}). Using Eq.~(\ref{eq:Nsmallsmall}) in (\ref{eq:dtilde}), the inverse Laplace transform reads
\begin{equation}
\tilde{D}(\nu) = \exp \left( -\alpha \sqrt{2\pi \nu\omega_c} \right).  
\end{equation}
which leads, {\it via}~(\ref{eq:laplace}), to~\cite{Baier:2001yt}
\begin{equation}\label{eq:BDMSestimate}
\Dbar(\epsb) = \alpha\,\sqrt{\frac{1}{2\,\epsb^3}}\,\exp\left(-\frac{\pi\alpha^2}{2\epsb}\right).
\end{equation}

The expression~(\ref{eq:BDMSestimate}) is plotted in Figure~\ref{fig:comparisonBDMS} ({\it dashed}) as a function of the energy loss $\epsb$, together with the full calculation for outgoing quarks ({\it solid}). Whereas the location of the peak of the distribution is roughly similar ($\epsb\approx 0.1$), it exhibits a much larger energy tail than what we found numerically\footnote{Remark for instance the infinite mean energy loss $\langle\eps\rangle$ in Eq.~(\ref{eq:BDMSestimate}).}. Consequently, the probability for a small energy loss is somewhat reduced as compared to our full result.

\begin{figure}[htbp]
\begin{center}
\includegraphics[width=9.cm]{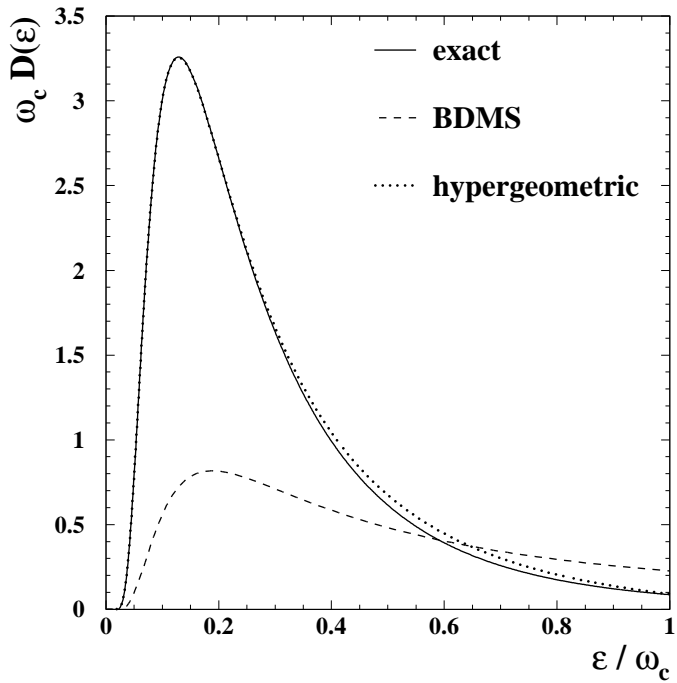}
\caption{Comparison of the distribution $\Dbar(\epsb)$ computed numerically for outgoing quarks with the BDMS analytic estimate (\ref{eq:BDMSestimate}) from the small $\ombar$ behavior (\ref{eq:Nsmallsmall}). Also shown in dotted line the analytic formula for $\Dbar(\epsb)$ extracted from the approximated gluon multiplicity~(\ref{eq:Nsmall_out}).}
\label{fig:comparisonBDMS}
\end{center}
\end{figure}

The poor agreement between the illustrative guess~(\ref{eq:BDMSestimate}) and our exact probability distribution could have been anticipated from Figure~\ref{fig:mult} where the small $\ombar$ approximation for $N(\omega)$ ({\it dash-dotted}) was shown to fail on the whole energy range. On the contrary, as stressed in the previous section, one might guess that the gluon multiplicity~(\ref{eq:Nsmall_out}) (respectively, (\ref{eq:Nsmall_in})) emitted by an outgoing (respectively, incoming) quark gives more satisfactory results. The inverse Laplace transform $\tilde{D}(\nu)$ now reads~\cite{Baier:2001yt}
\begin{equation}
 \tilde{D}(\nu) \label{eq:Dtapproxout}
\>\simeq\> \exp \left[ -\alpha\left(\sqrt{2\pi \nu\omega_c}  
- \ln2 \ln(\nu\omega_c) - 1.84146 \right)  \right] , 
\end{equation}
for an outgoing quark case, whereas we found
\begin{equation} \label{eq:Dtapproxin}
 \tilde{D}(\nu) 
\>\simeq\> \exp \left[ -\alpha\left(\sqrt{2\pi \nu\omega_c}  
- 0.981755\, \ln(\nu\omega_c) - \frac{1}{4}  \ln^2(\nu\omega_c) - 2.21561 \right)  \right] , 
\end{equation}
for incoming quarks. Therefore, the quenching weights $D^{out}(\eps)$ and $D^{in}(\eps)$ can be determined analytically through the inverse Laplace transforms of ({\ref{eq:Dtapproxout}) and ({\ref{eq:Dtapproxin}), respectively. 

Although analytic, the expressions obtained are lengthy and hardly transparent (sum of hypergeometrical functions), and are thus not reproduced here. Rather, we display in Figure~\ref{fig:comparisonBDMS} the excellent agreement between the analytic formula for $D^{out}(\eps)$ ({\it dotted}) together with the result computed from the exact gluon spectrum. We may notice in particular that the agreement remains perfect up to energy loss $\eps \approx 0.3\,\omega_c$, above which the small energy approximations ({\ref{eq:Nsmall_out}) and (\ref{eq:Nsmall_in}) for $N(\omega)$ start to fail (See Section~\ref{subse:mult} and Figure~\ref{fig:mult}).

\subsection{Analytical parameterization}\label{subse:param}

As mentioned in the introduction, the probability distribution bridges the gap between the theory of medium-induced parton energy loss on the one hand and the observable consequences on the other hand. In particular, its knowledge is required to model the quenching of hadron spectra in nuclear collisions. This was the main motivation for our present study. 

However, we have seen that $D(\eps)$ cannot be solved analytically, and thus neither can the  medium cross section~(\ref{eq:mediumxs}). Even though we have just stressed that an analytic expression is shown to mimic almost perfectly the numerical results, its complicated expression makes it useless in practical terms. For the sake of simplicity, we shall give instead in this section an empirical analytical expression for $D(\eps)$.  

The energy dependence of the quenching weight $D(\eps)$ follows with a great accuracy a log-normal distribution,
\begin{equation}\label{eq:lognormal}
\Dbar(\epsb) = \frac{1}{\sqrt{2\,\pi}\, \sigma\,\epsb}\,\exp\left[-\frac{\left(\log{\epsb}-\mu\right)^2}{2\,\sigma^2}\right]
\end{equation}
characterized by two parameters, $\mu$ and $\sigma$. The very nice agreement between the log-normal parameterization and the full result is illustrated in Figure~\ref{fig:fit} for outgoing ({\it left}) and incoming ({\it right}) quarks, using the parameters given in Table~\ref{tab:parameters}\footnote{Although we give this two parameter set, a small but clear anticorrelation exists between $\mu$ and $\sigma$.}. 
 \begin{table}[htbp]
\begin{center}
\begin{tabular}{|p{1.6cm}|p{1.6cm}|p{1.6cm}|}
\hline
\cline{2-3} 
 &\centerline{out} & \centerline{in} \\[-0.4cm]
\hline\\[-0.45cm]
\hline\\[-0.4cm]
\centerline{$\mu$} & \centerline{-1.5}  & \centerline{-2.55} \\[-0.4cm]
\hline\\[-0.4cm]
\centerline{$\sigma$} & \centerline{0.73} & \centerline{0.57} \\[-0.4cm]
\hline
\centerline{$\langle \eps \rangle$} & \centerline{0.3 $\omega_c$} & \centerline{0.1 $\omega_c$}\\
\hline
\end{tabular}
\end{center}
\caption{($\mu, \sigma$) parameters of the analytic approximation (\ref{eq:lognormal}) to the distribution $\Dbar(\epsb)$ for both outgoing ({\it left}) and incoming ({\it right}) quarks.}
\label{tab:parameters}
\end{table}
\begin{figure}[htbp]
\begin{center}
\includegraphics[width=17.5cm]{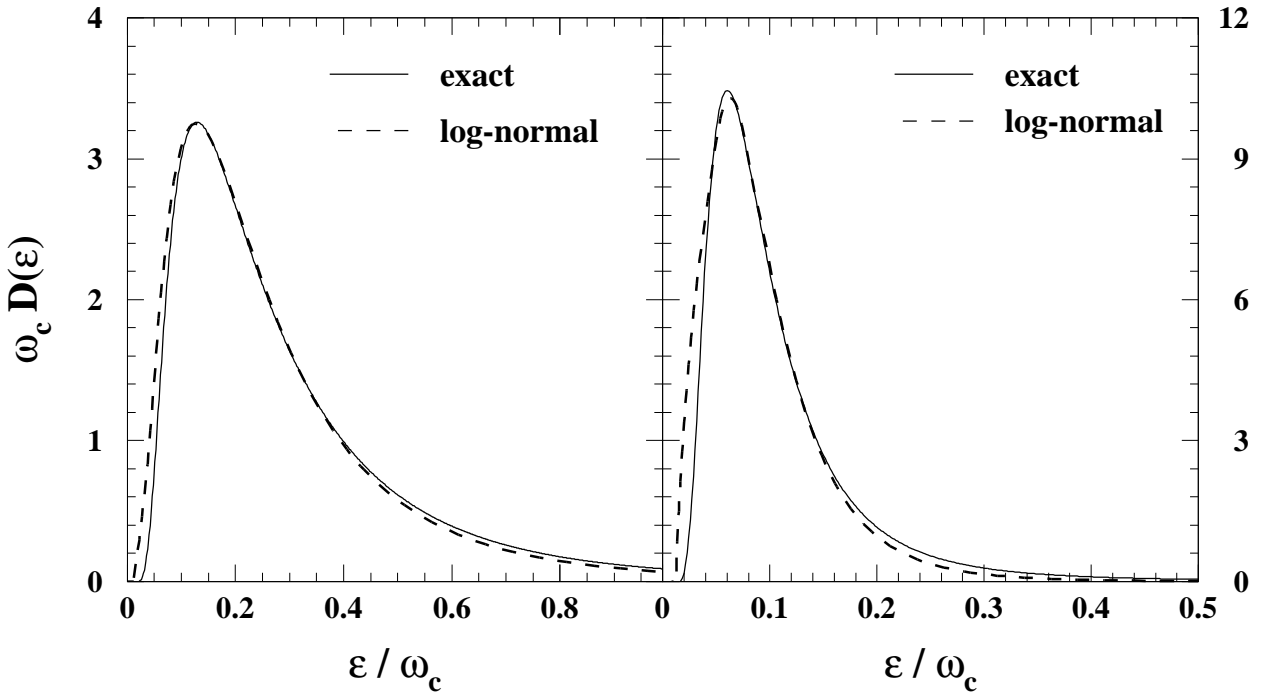}
\caption{Distribution $\Dbar(\epsb)$ ({\it solid}) for outgoing ({\it left}) and incoming ({\it right}) quarks together with the log-normal distributions Eq.~(\ref{eq:lognormal}) ({\it dashed}).}
\label{fig:fit}
\end{center}
\end{figure}

The simple analytic expression~(\ref{eq:lognormal}) can therefore easily be used to compute the medium cross section~(\ref{eq:mediumxs}) and hence the quenching of hadron yields.

To give the reader another feeling for the agreement between the parameterization~(\ref{eq:lognormal}) and the exact result, the mean energy loss experienced by the fast parton in the medium,
\begin{equation}
\langle \eps \rangle = \exp\left(\mu + \frac{1}{2}\,\sigma^2\right)\,\omega_c \, ,
\end{equation}
is computed in Table~\ref{tab:parameters}. The values tend to be pretty close to the result found analytically by BDMS~\cite{Baier:1998kq}
\begin{equation}
\begin{split}
\langle \eps \rangle = \frac{\alpha_s\,C_F}{2}\,\,\omega_c  = \frac{1}{3}\,\,\omega_c\, ,&\qquad \mathrm{(outgoing\,\,\,\, quark)}\\
\langle \eps \rangle = \frac{\alpha_s\,C_F}{6}\,\,\omega_c = \frac{1}{9}\,\,\omega_c\, .&\qquad \mathrm{(incoming\,\,\,\, quark)}
\end{split}
\end{equation}

BDMS~\cite{Baier:2001yt} emphasized that the standard modeling of the quenching using the {\it mean} energy loss 
\begin{equation}\label{eq:mediumxsbias}
\sigma^{medium} \simeq D(\langle\eps\rangle)\,\otimes\,\sigma^{vacuum}
\end{equation}
instead of (\ref{eq:mediumxs}) leads to a strong bias ---~when $\sigma^{vacuum}$ is a steeply falling function~--- whose strength is given by the higher moments of $D(\eps)$~\cite{Baier:2001yt}. In particular, the larger the asymmetry of the quenching weight (characterized by its skewness parameter $\gamma=\langle\epsb\,^3\rangle/\langle\epsb\,^2\rangle^{3/2}$), the stronger the bias effect. Indeed, we remark for instance in Figure~\ref{fig:fit} that the mean $\langle \eps \rangle$ proves larger than the most probable value of the distribution. In addition to that, let us note that this bias will become more pronounced when considering outgoing than incoming quarks, from the larger skewness of the distribution ($\gamma \approx 3.1$ and $2.1$, respectively).

Although certainly useful, one may nevertheless wonder about the origin of the log-normal dependence of the probability distribution $D(\eps)$. The reason for this is not clear. Even though it could be accidental, we argue it may rather come from the effect of the combinatorics in the Poisson approximation~(\ref{eq:poisson}). Let us be more explicit. Because the BDMPS medium-induced spectrum $dI/d\omega$ dramatically drops with the gluon energy $\omega$, the energy loss $\eps$ will be carried away by a large\footnote{Strictly speaking, the series has to be entirely resummed because of the vanishing term $\exp\left(-N(0)\right)$ in Eq.~(\ref{eq:poisson}).} number of soft radiated gluons. It could be the huge product of radiation probabilities in the Poisson approximation (\ref{eq:poisson}) that is responsible for such a behavior. We therefore conjecture that a log-normal quenching weight may emerge for any fast falling medium-induced gluon spectrum. This will be further discussed in the next section where the series~(\ref{eq:poisson}) is partially summed.

%%%%%%%%%%%%%%%%%%%%%%%%%%%%%%%%%%%%%%%%%%%%%%%%%%%%%%%%%%%%%%%%%%%%
\section{Approximations}\label{se:approx}
%%%%%%%%%%%%%%%%%%%%%%%%%%%%%%%%%%%%%%%%%%%%%%%%%%%%%%%%%%%%%%%%%%%%

\subsection{Bethe-Heitler regime}

The lifetime $t$ of the medium induced gluons emitted by the hard quark is given by~\cite{Baier:1995bd}
\begin{equation}
t = \frac{\omega}{k_\perp^2},
\end{equation}
where the transverse momentum $k_\perp^2$ of the gluon and its mean free path $\lambda_g$ are related through the transport coefficient $\hat{q} = \mu^2 / \lambda_g$~\cite{Baier:1998kq}. As long as the lifetime $t$ remains much smaller than the typical distance between two scattering centers in the medium, the radiation spectrum will be proportional to the gluon emission on one single radiator, that is
\begin{equation}\label{eq:BHspectrum}
\frac{dI(\omega)}{d\omega}^{BH} = \frac{L}{\lambda_q}\,\times\,\left(\frac{dI(\omega)}{d\omega}\right)_{(1)}
\end{equation}
where $L / \lambda_q$ is the number of collisions encountered by the leading quark with a mean free path $\lambda_q$. This is the Beithe-Heitler (BH) regime that occurs in QCD for small lifetime gluons.

On the contrary, a gluon with a long lifetime $t$ (as compared to its mean free path $\lambda_g$) will only see a group of scattering centers as a whole. In this regime, the induced gluon spectrum ({\ref{eq:BHspectrum}) is suppressed by the number of scattering centers $N_{coh}$ that act coherently~\cite{Baier:1995bd},
\begin{equation}\label{eq:LPMspectrum}
\frac{dI(\omega)}{d\omega}^{LPM} = \frac{L}{\lambda_q}\,\times \left( \frac{dI(\omega)}{d\omega} \right)_{(1)} \, \times \, \frac{1}{N_{coh}}
\end{equation}
which is the Landau-Pomeranchuk-Migdal (LPM) gluon spectrum~(\ref{eq:dIdo_out}) and (\ref{eq:dIdo_in}) computed by BDMPS. Hence, this coherent regime will set in for gluon energies greater than
\begin{equation}\label{eq:omega_min}
\omega_{min} = k_\perp^2\,\lambda_g \simeq \mu^2 \,\lambda_g.
\end{equation}
A rough estimate $\omega_{min} \approx 300$~MeV for a hot QCD medium is given by BDMS in~\cite{Baier:2001yt}. At small gluon energy, the induced gluon spectrum is negligible as compared that of the extrapolation of the LPM spectrum~(\ref{eq:dIdo_in}),
\begin{equation}
\omega\,\frac{dI(\omega)}{d\omega} = \mathrm{constant} \ll \omega\,\frac{dI(\omega)}{d\omega}^{LPM} \qquad \mathrm{at}\,\,\,\omega < \omega_{min}.
\end{equation}
As already pointed out at the end of the previous section, the quark energy loss actually originates from the emission of a large number of {\it soft} gluons, with energy $\omega \ll \epsilon \simeq \cO{\omega_c}$, because of the divergence of the BDMPS gluon spectrum in the infrared sector. Hence, we anticipate a large sensitivity in our results to the Bethe-Heitler regime ($\omega < \omega_{min}$). To estimate this sensitivity, we have repeated the computation of the quenching weight using a somewhat arbitrary truncated induced gluon spectrum,
\begin{equation}
\frac{dI(\omega < \omega_{min})}{d\omega} = 0
\end{equation}
where the IR cutoff $\omega_{min}$ is expressed as a function of the scale $\omega_c$.

As a consequence, the number of gluons radiated by the hard quark $N(0) = N(\omega_{min})$ becomes finite, unlike in Section~\ref{se:setup} where it was shown to be slowly divergent in the IR. Using (\ref{eq:omega_min}) in (\ref{eq:Nsmallsmall}), it is given by the number of collisions $L/\lambda_q$ (or opacity) in the medium~\cite{Baier:2001yt}. Therefore, the probability $p_0$ for having no interaction, 
\begin{equation}
p_0 = \exp \left[ - \int_0^{+\infty} d\omega \, \frac{dI(\omega)}{d\omega} \right] \approx \exp\left(-L/\lambda_q\right)
\end{equation}
that appears in Eq.~(\ref{eq:poisson}) no longer vanishes in a finite length medium. In the general case, the probability distribution will read~\cite{W,Salgado:2002cd}
\begin{equation}\label{eq:Dgeneral}
D(\eps) = p_0\,\delta(\eps) + d(\eps).
\end{equation}
Following Wiedemann~\cite{W}, we shall subtract the discrete contribution in (\ref{eq:Dgeneral}) before performing the numerical inverse Laplace transform,
\begin{equation}
d(\eps) = \int_{\cal{C}} \, \frac{d\nu}{2\pi i} \>
(\tilde{D}(\nu)-p_0)\,e^{\nu {\eps}}
\end{equation}
to eventually extract the continuous part $d(\eps)$ of the probability distribution.
\begin{figure}[htbp]
\begin{center}
\includegraphics[width=9.cm]{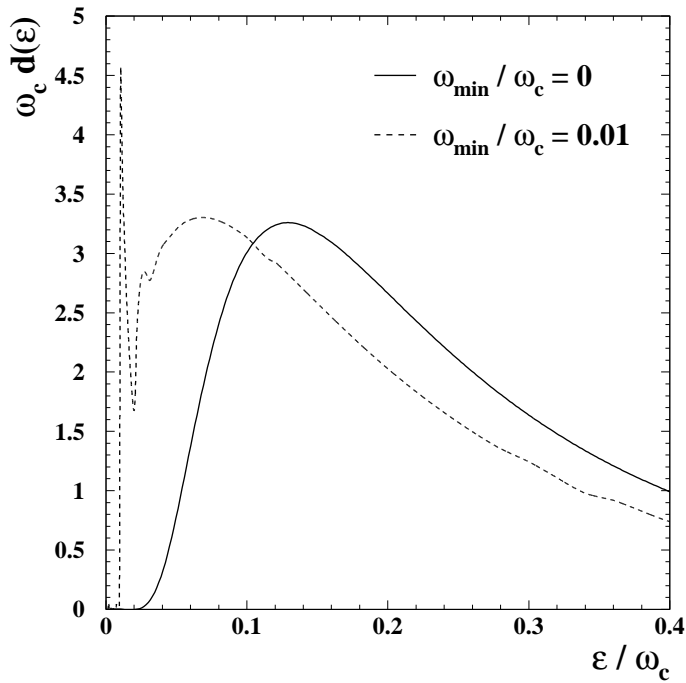}
\caption{Distribution $\bar{d}(\epsb)$ for outgoing quarks computed from the BDMPS spectrum with ({\it dashed}) and without ({\it solid}) IR cutoff $\omega_{min}$.}
\label{fig:bh}
\end{center}
\end{figure}

The result for $d(\eps)$ assuming $\omega_{min} = \omega_c / 100$ is shown in Figure~\ref{fig:bh} ({\it dashed}) together with the previously computed $D(\eps)$ without any IR cutoff ({\it solid}). Taking $\hat{q}\simeq 5$~GeV/fm$^2$ and $L = 5$~fm in~(\ref{eq:omc}), this would correspond to $\omega_{min} \simeq 600$~MeV. First, Figure~\ref{fig:bh} clearly indicates that the shape of the distribution looks pretty similar to what has been obtained before, although shifted to smaller energy loss. This does not come as a surprise as fewer gluons are radiated. Another remarkable feature is the structure observed at $\eps \ll \omega_c$. This actually originates from the emission of a very small number of gluons $n$ in the Poisson series~(\ref{eq:poisson}). To go a bit further, we plot in Figure~\ref{fig:bhzoom} the quenching weight as a function of $\eps$ in units of the cutoff $\omega_{min}$. The distribution ({\it solid line}) is identically equal to zero up to $\epsilon = \omega_{min}$ when the channel for the one gluon emission opens, followed by a strong decrease coming from the dropping gluon spectrum~(\ref{eq:dIdo_out}). Angular points may clearly be seen at $\epsilon / \omega_{min} = 2$ and 3 which correspond to the opening of the two and three gluon radiation channels, respectively.
\begin{figure}[htbp]
\begin{center}
\includegraphics[width=9.cm]{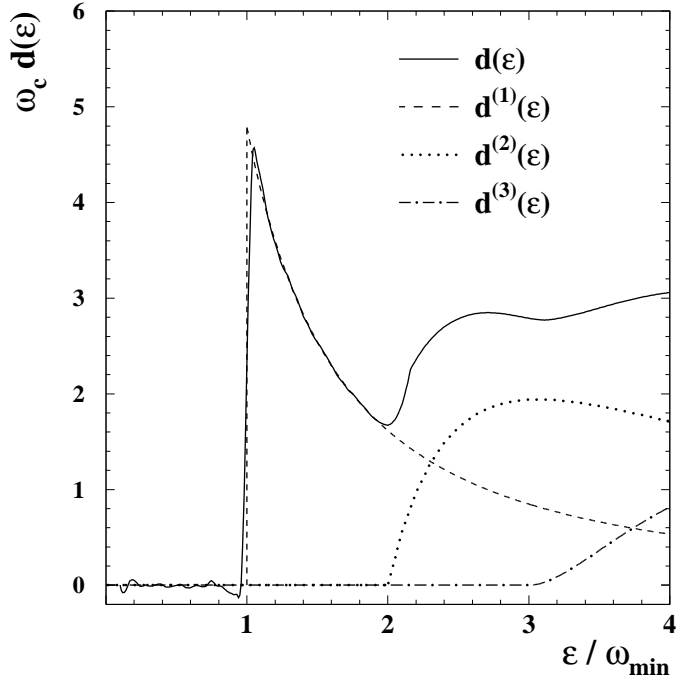}
\caption{Distribution $\bar{d}(\epsb)$ for outgoing quarks computed from the BDMPS spectrum with an IR cutoff $\omega_{min} = \omega_c /100$ ({\it solid}) together with the probability distributions $\omega_c\, d^{(i)}(\eps)$ to have $i = 1$~({\it dashed}), 2  ({\it dotted}), 3 ({\it dash-dotted}) gluons emitted (see text for details).}
\label{fig:bhzoom}
\end{center}
\end{figure}

Because each term is finite in the Poisson expression (recall, $p_0 > 0$), it becomes possible to sum all the individual contributions to reconstruct the quenching weight, as
\begin{equation}\label{eq:partial_sum}
D(\eps < m\,\omega_{min}) = p_0 \,\delta(\eps) + \sum_{n = 1}^{m-1} d^{(n)}(\eps),
\end{equation} 
where the $d^{(n)}$ represent the probability for having exactly $n$ gluons emitted and where the terms $n \ge m \approx \eps/\omega_{min}$ do not contribute. Although the resummation procedure offered by BDMS allows one to compute elegantly all these terms, we nevertheless compute the first three terms of Eq.~(\ref{eq:partial_sum}), which can be written as
\begin{equation}
d^{(1)}(\epsilon) = p_0\,\times\,\frac{dI(\epsilon)}{d\omega}\,\times\,\Theta(\epsilon-\omega_{min}),
\end{equation}
\begin{equation}
d^{(2)}(\epsilon) = \frac{p_0}{2}\,\times\,\int_{\omega_{min}}^{\epsilon-\omega_{min}} d\omega_1 \,\frac{dI(\omega_1)}{d\omega}\,\frac{dI(\epsilon-\omega_1)}{d\omega}\,\times\,\Theta(\epsilon-2\,\omega_{min}),
\end{equation}
and,
\begin{equation}
d^{(3)}(\epsilon) = \frac{p_0}{6}\,\times\,\int_{\omega_{min}}^{\epsilon-\omega_{min}} d\omega_1 \,\frac{dI(\omega_1)}{d\omega}\,\int_{\omega_{min}}^{\epsilon-\omega_1-\omega_{min}} d\omega_2\,\frac{dI(\omega_2)}{d\omega}\,\frac{dI(\epsilon-\omega_1-\omega_2)}{d\omega}\,\times\,\Theta(\epsilon-3\,\omega_{min}),
\end{equation}
for the emission of 1, 2, and 3 gluons. These are displayed in Figure~{\ref{fig:bhzoom} as dashed, dotted, dash-dotted lines, respectively. Their sum is shown to reproduce exactly the full result in solid line. At larger energy loss $\eps \gg \omega_{min}$, the sum over the large number $m \approx \eps / \omega_{min} \gg 1$ of gluons makes the distribution $d(\eps)$ a smooth function of the energy $\eps$. We then recover the log-normal behavior previously discussed.

\subsection{Beyond the soft gluon approximation}

The calculations performed so far have been computed in the {\it soft} gluon approximation, i.e., assuming the gluon energy $\omega$ to remain small with respect to the leading quark energy $E$. This approximation is, however, rarely justified in practice. BDMS give in Ref.~\cite{Baier:1998kq} the full expression of the induced spectrum radiated by a quark. Neglecting $\cO{(\omega/E)^2}$ terms, it can be written as 
\begin{equation}\label{eq:spec_E}
\frac{dI(\omega, E)}{d\omega} = (1-\frac{\omega}{E})\,\times\,\frac{dI(\omega)}{d\omega}\,\times\,\Theta(E - \omega)
\end{equation}
where the additional factor actually comes from the quark-gluon DGLAP splitting function.

The probability distribution $D(\eps,E)$ is represented in Figure~\ref{fig:largex} for various quark energies $E/\omega_c$. The effect of the $\cO{\omega/E}$ corrections in the gluon spectrum~(\ref{eq:spec_E}) is to reduce hard gluon emission, and hence the high-energy tail of the distribution. On the contrary, the small $\eps$ behavior of the quenching weight remains unchanged, with the exception of the absolute magnitude which follows from the normalization constraint. In particular, the location of the peak does not exhibit a strong quark energy dependence. Perhaps more interesting is the following observation. Although Eq.~(\ref{eq:spec_E}) ensures that a single gluon cannot carry more energy than available ($\omega < E$), the use of the spectrum~(\ref{eq:spec_E}) does not a priori guarantee the quark total energy loss 
\begin{equation}
\eps = \sum_i \, \omega_i
\end{equation}
in the Poisson expression~(\ref{eq:poisson}) to be bounded. Indeed, we do observe in Figure~\ref{fig:largex} a small but significant contribution of the probability distribution in this kinematically forbidden region, $D(\eps > E) \ne 0$, when the quark energy $E$ is small enough. This signals the breakdown of the eikonal approximation on which the BDMPS framework rely. It is indeed no longer justified to consider multiple successive and independent quark-nucleon scatterings when the quark energy is smaller than, say, half the energy scale $\omega_c$.
\begin{figure}[htbp]
\begin{center}
\includegraphics[width=9.cm]{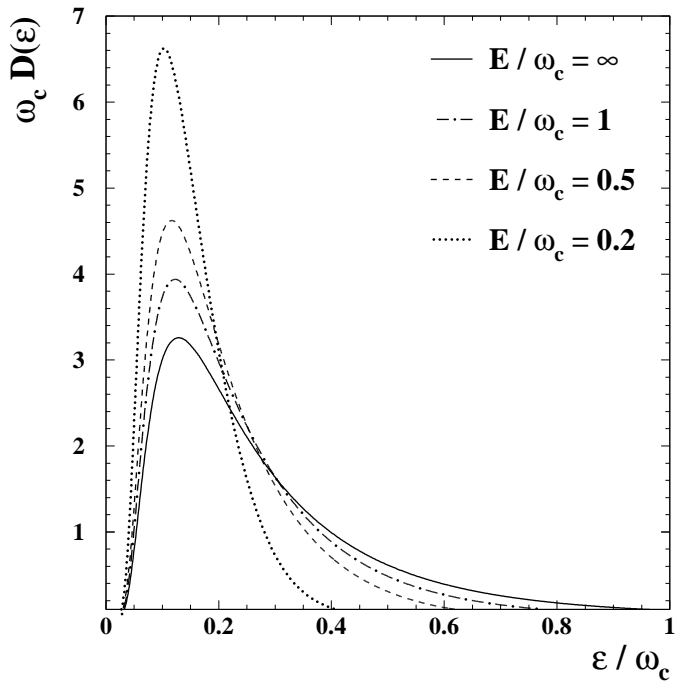}
\caption{Outgoing quark energy dependence of the distribution $\Dbar(\epsb,\Et)$ computed from the BDMPS spectrum Eq.~(\ref{eq:spec_E}). Shown in solid line is the previous result obtained in the soft limit.}
\label{fig:largex}
\end{center}
\end{figure}

The need for a simple analytic parameterization of the probability distribution to compute the quenching of hadronic spectra has been stressed in Section~\ref{subse:param}. Hence, we shall now extend this parameterization for any quark energy. Noticing that the distributions $D(\eps,E)$ still follow log-normal distributions, the quark energy dependence will enter through the parameters $\mu$ and $\sigma$. The quenching weight thus reads
\begin{equation}\label{eq:lognormal_E}
\Dbar(\epsb, \Et) = \frac{1}{\sqrt{2\,\pi}\, \sigma(\Et)\,\epsb}\,\exp\left[-\frac{\left(\log{\epsb}-\mu(\Et)\right)^2}{2\,\sigma(\Et)^2}\right]
\end{equation}
where the $\mu(E)$ and $\sigma(E)$ are given by the empirical laws
\begin{equation}
\begin{split}
\mu(\Et) & = -1.5 + 0.81 \times \left( \exp \left( -0.2/ \Et \right) -1 \right),\\
\sigma(\Et) & = \,\,0.72 + 0.33 \times \left(\exp\left(-0.2/ \Et \right) -1 \right).
\end{split}
\end{equation}

\begin{figure}[htbp]
\begin{center}
\includegraphics[width=9.cm]{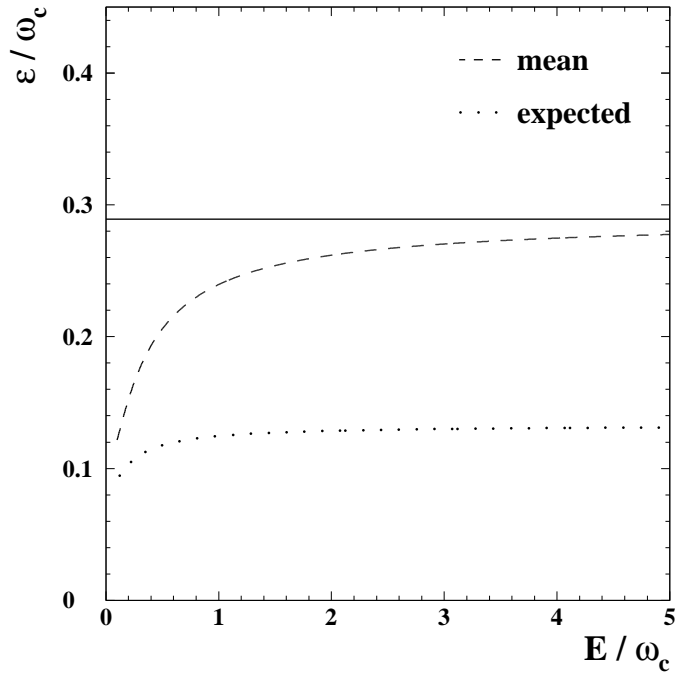}
\caption{Mean $\langle \eps \rangle$ ({\it dashed}) versus expected ({\it dotted}) energy loss as a function of the outgoing quark energy $E$. The solid line represents the mean energy loss determined in the soft limit.}
\label{fig:mean}
\end{center}
\end{figure}

Using the parameterization~({\ref{eq:lognormal_E}), the mean energy loss $\langle \eps \rangle$ is plotted as a function of the quark energy ({\it dashed}) in Figure~\ref{fig:mean}. In Ref.~\cite{Baier:1997kr}, BDMPS discussed in a schematic way the characteristics of the $E$ dependence of the mean energy loss $\langle \eps \rangle$. They found that~\cite{Baier:1997kr}
\begin{equation}\label{eq:schematic}
\langle \eps \rangle\bigm/\omega_c \propto 
\left\{
\begin{array}{rl}
(E/\omega_c)^{1/2} & \qquad {\mathrm{if}}\ E < \omega_c\\
1 & \qquad {\mathrm{if}}\ E > \omega_c\\
\end{array}
\right.
\end{equation}
These features are qualitatively reproduced here. At high energy $E \gg \omega_c$, $\langle \eps \rangle$ is independent of $E$ and rather close to its asymptotic value $\langle \eps \rangle \simeq 0.3\, \omega_c$ (shown in solid line), whereas a stronger dependence is seen at low energy, $E \sim \omega_c$. In particular, it is interesting to note that the $L^2$ dependence of $\langle \eps \rangle$ will not set in until approximately $E \simeq 3\,\omega_c$. Our result is also reminiscent of what has recently been found by Gyulassy, L\'evai, and Vitev in the computation of the mean multiplicity of radiated gluons~\cite{Gyulassy:2001nm}, that strongly increases up to a quark energy $E \approx \omega_c \approx 15$~GeV in their calculation. Finally, we show in Figure~\ref{fig:mean} ({\it dotted}) the expected energy loss (that we define as the most probable value of the quenching weight) that is well smaller than the mean $\langle \eps \rangle$ and independent of the quark energy.

%%%%%%%%%%%%%%%%%%%%%%%%%%%%%%%%%%%%%%%%%%%%%%%%%%%%%%%%%%%%%%%%%%%%
\section{Applications}\label{se:app}
%%%%%%%%%%%%%%%%%%%%%%%%%%%%%%%%%%%%%%%%%%%%%%%%%%%%%%%%%%%%%%%%%%%%

The probability distributions for both incoming and outgoing quarks have been computed in the previous sections. To illustrate the use of these results, we now determine the quenching of hadron spectra in nuclear collisions and compare it to experimental preliminary data.

\subsection{Medium modification of fragmentation functions}

The HERMES collaboration at DESY recently reported on hadron yields measured in electron-nucleus collisions. They measured the production ratio
\begin{equation}\label{eq:suppDIS}
R_A^{h}(z,\nu) = \frac{1}{N_A^e(\nu)}\,\frac{N_A^h(z,\nu)}{d\nu\,dz}\Biggm/\frac{1}{N_D^e(\nu)}\,\frac{N_D^h(z,\nu)}{d\nu\,dz}
\end{equation}
in a ``heavy'' (N and Kr) over a light (D) nucleus for a given hadron species $h$. Here, $\nu$ denotes the virtual photon energy in the lab frame, $z$ the momentum fraction carried by the produced hadron, and where the multiplicity of produced electrons $N_A^e$ normalizes the hadron yield $N_A^h$.

The hadron multiplicity in~(\ref{eq:suppDIS}) can be computed perturbatively to leading order (LO) in $\alpha_s$. It is written in terms of (nuclear) parton densities $q_f(x, Q^2,A)$ and fragmentation functions (FF) $D_q^h(z, Q^2,A)$,~\footnote{Isospin corrections should be small as we compare nuclei with a similar $Z/A$ ratio, and have thus been neglected.}
\begin{eqnarray}\label{eq:multDIS}
\frac{1}{N_A^e}\frac{dN_A^h}{d\nu\,dz} & = & \int \, dx\, \sum_f \, e^2_f\, q_f(x, Q^2, A) \, \sigma^{\gamma^* q}(x,\nu) \, D_f^h(z, Q^2, A) \biggm/ \\
& & \qquad \qquad \qquad \qquad \qquad \qquad \qquad \int \, dx\, \sum_f \, e^2_f\, q_f(x, Q^2, A) \, \sigma^{\gamma^* q}(x,\nu) \nonumber
\end{eqnarray}
and where the LO $\gamma^* q$ cross section is given by
\begin{equation}
\sigma^{\gamma^* q}(x,\nu) = \frac{4\pi \alpha_s^2(Q^2)\,M \,x}{Q^4} \times \left[ 1+\left(1-\frac{Q^2}{x\,s}\right)^2 \right].
\end{equation}
The integral over Bjorken $x = Q^2 / (2M\nu)$ appearing in~(\ref{eq:multDIS}) is given by the $Q^2$ acceptance of the HERMES experiment. To a first approximation, only the valence up quark will contribute to the hadron yield~(\ref{eq:multDIS}) when $x$ is not too small. Hence, the ratio~(\ref{eq:suppDIS}) will approximately be given by the ratio of the $u\to h$ fragmentation functions
\begin{equation}
R_A^{h}(z,\nu) \simeq D_u^h(z, Q^2, A) \biggm/ D_u^h(z, Q^2, D).
\end{equation}
Therefore, the nuclear dependence of the fragmentation functions might be revealed through the measure of $R^h$. We further note that the effects of nuclear shadowing in the parton densities $q_f(x, Q^2, A) $ should remain small as they mainly cancel in the ratio $N^h / N^e$~(\ref{eq:multDIS}). 

\begin{figure}[htbp]
\begin{center}
\includegraphics[width=14.5cm]{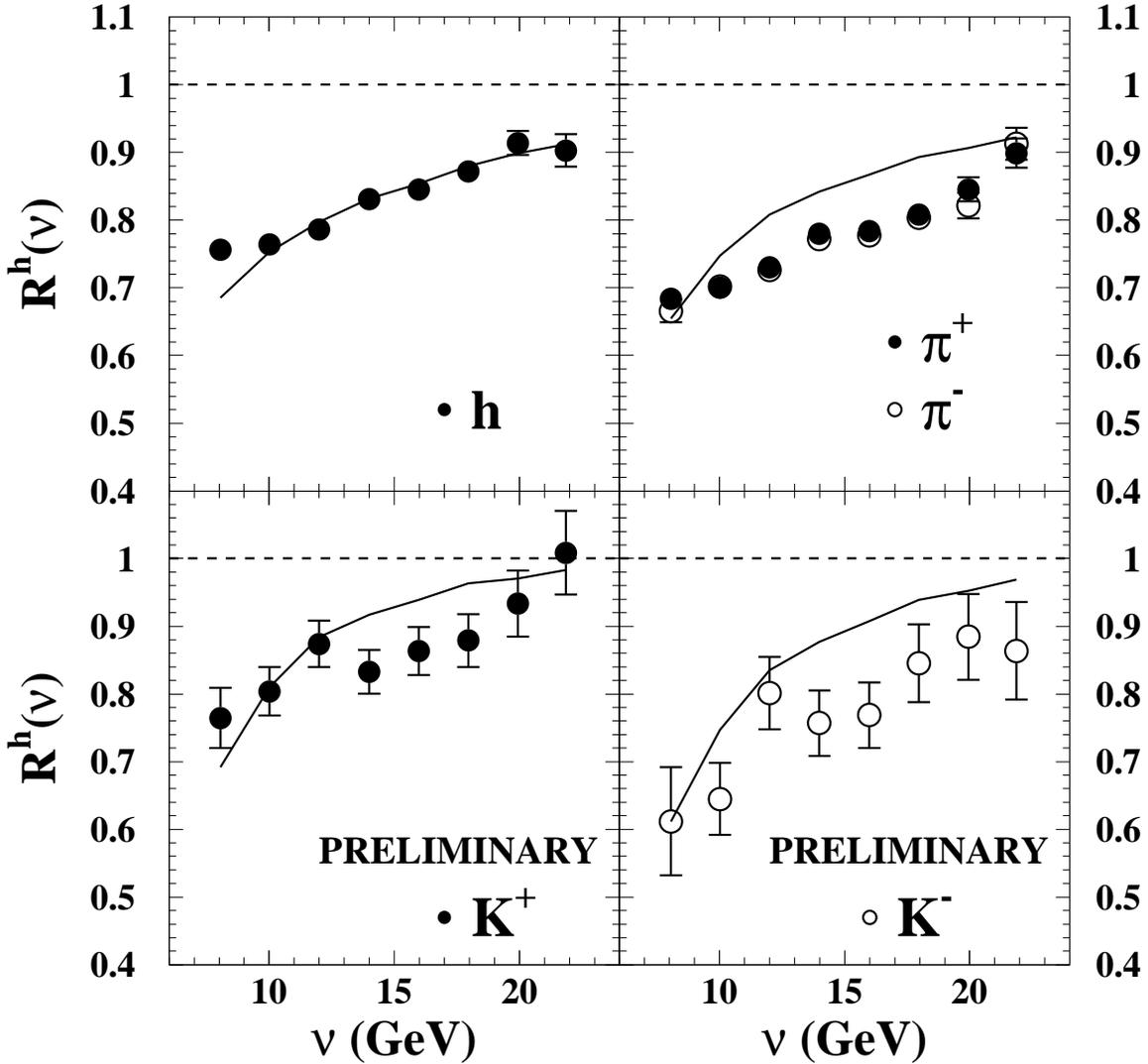}
\caption{Attenuation ratio $R_{Kr}^h(\nu)$ (\ref{eq:suppDIS}) plotted as a function of $\nu$ for charged hadrons (upper left), charged pions (upper right), positive and negative (lower) kaons. Calculations ({\it solid}) are compared to HERMES preliminary data ({\it circles}) taken from Ref.~\cite{A:2002ksM:2001znM:2002ea}.}
\label{fig:HERMES}
\end{center}
\end{figure}

The multiple scattering of the produced quark in the nuclear medium may be responsible for the observed dependence of the ratio $R_A^{h}$ with $\nu$ and $z$~\cite{Wang:2002ri}. Indeed, the energy loss of the hard quark will shift its energy from $E_q \simeq \nu$ to $E_q \simeq \nu - \eps$ at the time of the hadronization. Therefore, this mechanism leads to a shift in $z$,
\begin{equation}
z = \frac{E_h}{\nu} \qquad \to \qquad z^* = \frac{E_h}{\nu - \eps} = \frac{z}{1 - \eps/\nu}.
\end{equation}
The effect of the energy loss mechanism on the nuclear fragmentation functions may be modeled according to~\cite{Wang:1996yh,Wang:2002ri}
\begin{equation}\label{eq:modelFF}
z\,D_f^h(z, Q^2, A) = \int_0^{\nu - E_h} \, d\eps \,\,D(\eps, \nu)\,\,\, z^*\,D_f^h(z^*, Q^2).
\end{equation}

Using (\ref{eq:modelFF}) in (\ref{eq:multDIS}) and integrating over $z$, the $\nu$ dependence of $R^h(\nu)$ in a krypton over a deuterium target has been computed. In the calculation, the energy scale $\omega_c$ was determined by the transport coefficient $\hat{q}$ adjusted to the data and the length of matter covered by the hard quark proportional to the nuclear radius $R$, $L = 3/4\,R$. The parton densities were given by GRV98 LO~\cite{Gluck:1998xa} while we made use of the Kretzer LO parameterization for the fragmentation functions~\cite{Kretzer:2000yf}.

The calculations for charged hadrons, pions, and kaons are compared with HERMES preliminary data in Figure~\ref{fig:HERMES}. The trend is reproduced well for all hadron species, although the calculation for the pions ($\pi^+ + \pi^-)$ somehow underpredicts the effect. It is also interesting to notice that the $K^-$ yield is more suppressed than the $K^+$ ---~as seen in the data~--- which arises from the stronger slope of the FF at large $z$. A much smaller difference between the $\pi^+$ and $\pi^-$ attenuation (average in the Figure) is also observed, unlike the present data which do not exhibit any isospin dependence in this channel. Let us also mention that the $z$ dependence of the ratio is fairly reproduced when $z$ is not too large, the calculations predicting a much stronger suppression at large $z$ than what is observed experimentally. However, the formula (\ref{eq:modelFF}) we use may no longer be valid in this specific kinematic region where large higher twist corrections come into play~\cite{Wang:2002ri}.

The calculations were performed assuming the transport coefficient for nuclear matter to be~$\hat{q} = 0.75$~GeV/fm~$^2$. This would correspond to a mean energy loss per unit length $-dE/dz = \langle \eps \rangle / L$ to be $-dE/dz = 0.62$~GeV/fm in a large ($L\approx 5$~fm) nucleus. This result is close to what Wang and Wang determined in their analysis~\cite{Wang:2002ri}, although they had not considered the full probability distribution (but its mean) and neglect the $\nu$ dependence of the mean energy loss (which is found to be relevant at low $\nu \approx 8-10$~GeV). Furthermore, let us remark that this estimate is reduced by a factor of three for incoming quarks, $-dE/dz \approx 0.21$~GeV/fm, in excellent agreement with our recent Drell-Yan data analysis~\cite{Arleo:2002ph}.

It is not our aim to claim that energy loss is the only mechanism responsible for the trend observed in HERMES data. Indeed, many other effects such as nuclear absorption~\cite{B:1983knB:1987cf}, gluon bremsstrahlung~\cite{Kopeliovich:1995jt}, or partial deconfinement~\cite{N:1984pyJ:1984zwC:1985znA:2002yn} have been advanced to account for these measurements. Hence, the value extracted for $\hat{q}$ has to be seen as an upper limit.

\subsection{Quenching of high $p_\perp$ particles}

A huge parton energy loss is a clear signal for quark-gluon plasma (QGP) formation. Therefore, the production of high $p_\perp$ hadrons might reveal the existence, and more importantly the characteristics, of the produced medium in heavy ion collisions. Interpreting the depletion of the hadron yield observed at RHIC as coming from the energy loss experienced by hard quarks in an expanding QGP, we determine and compare the quenching factor
\begin{equation}\label{eq:qfactor}
R_{AA} (p_\perp) = \frac{d\sigma^{A A}(p_\perp)}{d p_\perp} \,\Biggm/\, A^2\,\frac{d\sigma^{p p}(p_\perp)}{d p_\perp}
\end{equation}
to PHENIX preliminary data on $\pi^0$ production in central Au-Au collisions at $\sqrt{s_{NN}} = 200$~GeV. The quenching factor~(\ref{eq:qfactor}) may be determined through the ratio\footnote{We neglect the longitudinal momentum of the produced quark ($p \approx p_\perp$).}~\cite{Baier:2001yt}
\begin{equation}\label{eq:quench}
R_{AA}(p_\perp) \approx \int_0^{+\infty}\,d\eps\,D(\eps,\, p_\perp + \eps)\,\times\,\frac{d\sigma^{vacuum}(p_\perp + \eps)}{d p^2_\perp} \quad\biggm/ \frac{d\sigma^{vacuum}(p_\perp)}{d p^2_\perp}
\end{equation}
where the $p_\perp$ differential vacuum cross sections can be computed perturbatively. Following BDMS, we shall however adopt the fit proposed by the PHENIX collaboration
\begin{equation}\label{eq:xsPHENIX}
   \frac{d\sigma^{{\rm vacuum}}(p_\perp)}{dp_\perp^2} \>\propto\>
\left(1.71 \>+\> p_\perp \,[{\rm GeV}]\right)^{-12.44}
\end{equation}
for the sake of simplicity. To determine~(\ref{eq:qfactor}), we use our parameterization~(\ref{eq:lognormal_E}) for $D(\eps,\, p_\perp)$, the energy scale $\omega_c$ being given by the dynamical scaling law of Salgado and Wiedemann~\cite{Salgado:2002cd}
\begin{equation}\label{eq:SW}
\langle \omega_c \rangle = \hat{q}(t_0)\,\int_{t_0}^{t_0 + L}\,dt' \,\left(t' - t_0 \right) \times \left(\frac{t_0}{t'}\right)^\alpha.
\end{equation}
where $\hat{q}(t_0)$ is the transport coefficient of the medium at the initial time $t_0$ and $\alpha=1$ characterizes the longitudinal expansion of the QGP. We take in the following an initial time $t_0 = 1$~fm, a medium length $L = 5$~fm and the transport coefficient $\hat{q}(t_0) = 3.5$~GeV/fm$^2$ (which leads to $\langle \omega_c \rangle \approx 10$~GeV).
\begin{figure}[htbp]
\begin{center}
\includegraphics[width=17.cm]{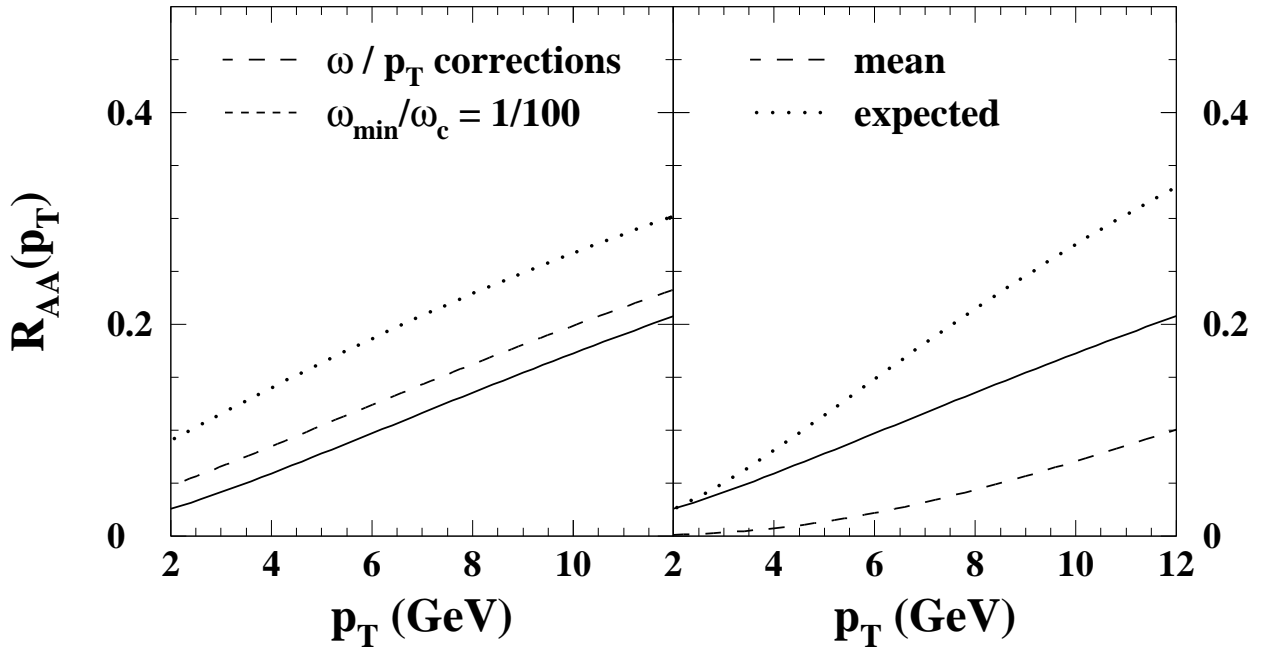}
\caption{{\it left:} The quenching factor $R_{AA}(p_\perp)$ is computed from the BDMPS gluon spectrum in the soft limit with ({\it dotted}) and without ({\it solid}) IR gluon energy cutoff $\omega_{min}$. Effects of $\cO{\omega/p_\perp}$ corrections to the soft gluon spectrum are shown in dashed line. {\it right :} The influence on the modeling of the quenching in the $p_\perp$ dependence of the quenching factor is displayed. The solid line represents the calculation from the full probability distribution~(\ref{eq:quench}), while the quenching factor is computed assuming a shift of the mean ({\it dashed}) and the expected ({\it dotted}) energy loss in the cross sections~(\ref{eq:xsPHENIX}).}
\label{fig:phenix2}
\end{center}
\end{figure}

The quenching~(\ref{eq:quench}) is plotted as a function of $p_\perp$ in the soft gluon approximation $p_\perp \gg \langle\omega_c\rangle$ in Figure~\ref{fig:phenix2} ({\it solid}). To give an idea of the uncertainties, we show on the left panel the effect of the $\cO{\omega/p_\perp}$ corrections in the induced spectrum~(\ref{eq:spec_E}) ({\it dashed}), that slightly ($\sim 10\%$) reduce the depletion. The ratio $R_{AA}(p_\perp)$ is also calculated assuming a gluon energy cutoff $\omega_{min} \approx 100$~MeV ({\it dotted}). The right panel displays the influence of the modeling of the quenching. The suppression is computed by shifting the mean ({\it dashed}) and the expected ({\it dotted}) energy loss in the cross section~(\ref{eq:xsPHENIX}) together with the result assuming the full distribution~({\it solid}). The effect of the bias is clear. The suppression is overestimated by more than a factor of two when one considers a mean energy loss rather than the full quenching weight. On the contrary, we note that shifting the cross sections by the peak value in the distribution underestimates the quenching.

The quenching~(\ref{eq:quench}) is now compared to the PHENIX $\pi^0$ preliminary measurements~\cite{d'Enterria:2002bw} in Figure~\ref{fig:phenix}. For this illustration, the transport coefficient $\hat{q}(t_0)= $~3.5~GeV/fm$^2$ was chosen so as to reproduce high $p_\perp$ data\footnote{In this calculation, we made use of the parameterization~(\ref{eq:lognormal_E}) for $D(\eps,p_\perp)$, i.e., with $\cO{\omega / p_\perp}$ corrections and a vanishing gluon energy cutoff $\omega_{min} = 0$. This corresponds to the dashed line in the left panel of Figure~\ref{fig:phenix2}.}. Because of the strong correlation existing between $\hat{q}$ and $L$ for a given $\langle\omega_c\rangle$, this absolute value should however be taken with great care. Figure~\ref{fig:phenix} clearly demonstrates that the trend of our estimate is opposite to what is observed experimentally, with the possible exception of the highest $p_\perp$ data bins. It is not obvious to see where this discrepancy comes from. First, the extrapolation to low $p_\perp$ is somehow difficult, as our results shall be much more sensitive to the IR behavior of the BDMPS spectrum~(\ref{eq:dIdo_out}) where the LPM regime is no longer guaranteed. Secondly, we have already stressed that the eikonal approximation~(\ref{eq:poisson}) breaks down for quark energies $p_\perp = \cO{\langle\omega_c\rangle/2}\simeq 5$~GeV. Moreover, other mechanisms may compete and weaken the quenching in this low $p_\perp$ region, such as the detailed balance process where the leading quark picks up thermally equilibrated gluons in the QGP~\cite{Wang:2001cs}. Let us finally mention that Vitev and Gyulassy recently suggested that the interplay of $k_\perp$ broadening, nuclear shadowing together with parton energy loss may account for the trend of the data~\cite{Vitev:2002pf}.

It has been assumed so far that the hard quark propagates through the QGP with length $L$ and subsequently hadronizes outside the medium. This picture should be true at high $p_\perp$ when the hadronization time is large enough because of Lorentz dilation. In the following, let us suppose that, in the $p_\perp$ range of interest here, the hard quark hadronizes {\it inside} the medium. The reader may of course worry about the relevance of the hadronization concept in a QGP. Thus, we rather imagine that hadronization might occur (and makes sense) in a cooling and more dilute ($t \gg t_0$) system such as a hot pion gas, although with a significant transport coefficient~\cite{Baier:2002tc}. The length covered by the parton is then no longer given by the system size $L$, but its hadronization time $t_h$,
\begin{equation}\label{eq:th}
t_h \simeq K(z)\,\times\,\frac{p_\perp}{\sigma}
\end{equation}
where $\sigma \simeq 1$~GeV/fm is the string tension. Several models have been proposed in the literature to characterize the $z$ dependence of the hadronization time~\cite{B:1983knB:1987cf,Kopeliovich:1995jt,Brodsky:1988xz}. Here, we take $K=1/2$ assuming a mean $\langle z\rangle =1/2$ either in the approach by Brodsky and Mueller~\cite{Brodsky:1988xz} or in the gluon bremsstrahlung model~\cite{Kopeliovich:1995jt}. In other words, the length $\sim t_h$ over which the quark propagates in the medium actually depends on (and increases with) $p_\perp$. We may write the mean $\langle \omega_c \rangle$ in this simplified picture as
\begin{equation}\label{eq:SWexp}
\langle \omega_c \rangle (p_\perp) = \hat{q}(t_0)\,\int_{t_0}^{t_0 + t_h(p_\perp)}\,dt' \,\left(t' - t_0 \right) \times \left(\frac{t_0}{t'}\right)^\alpha \qquad {\mathrm{if}} \qquad t_h(p_\perp) \lesssim L 
\end{equation}
The formula~(\ref{eq:SW}) should be replaced by~(\ref{eq:SWexp}) which depends approximately linearly with $p_\perp$ and thus so does the energy loss. As can be seen in Figure~\ref{fig:phenix}, this leads to a decrease of the quenching as a function of $p_\perp$ (as long as $t_h(p_\perp)\simeq\cO{L}$ over which Eq.~(\ref{eq:SWexp}) ceases to be valid) in good agreement with the trend of the data assuming $\hat{q}(t_0)= 4.5$~GeV/fm$^2$~\footnote{The quenching will essentially depend on the {\it product} $K\times\hat{q}(t_0)$, both terms being poorly known. It is needless to repeat that this transport coefficient chosen to account for the data is thus a rough estimate. This prevents us from drawing any conclusion from its absolute value.}. In this calculation, we have not considered possible final state interactions of the produced pion in the medium, which may somehow modify this picture. Provided these to remain rather small in a dilute system, a decrease followed by a subsequent increase of the quenching at larger $p_\perp$ would therefore be a signal of the transition between hadronization inside and outside of the hot medium. 

\begin{figure}[htbp]
\begin{center}
\includegraphics[width=9.cm]{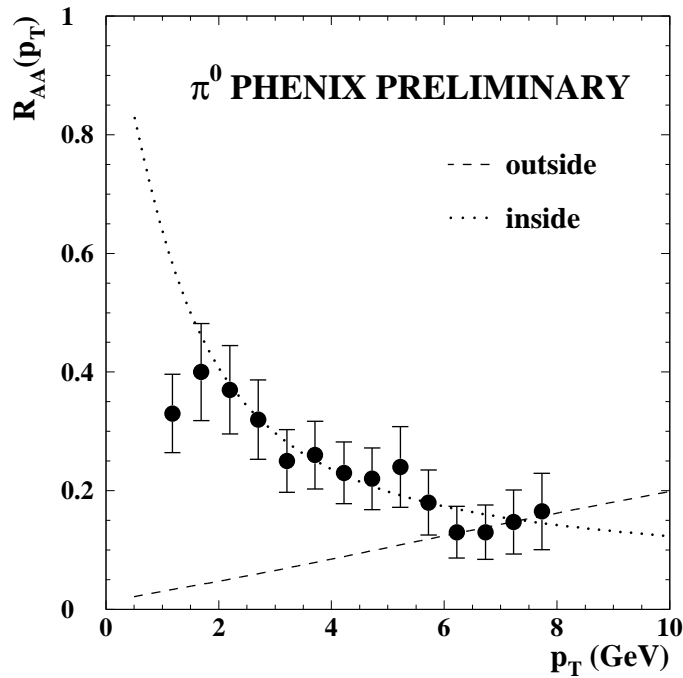}
\caption{Preliminary PHENIX data on $\pi^0$ suppression factor in Au-Au collisions at $\sqrt{s}=200$~GeV as a function of $p_\perp$ ({\it circles}) compared to the theoretical calculations assuming hadronization inside ({\it dotted}) and outside ({\it dashed}) the hot QCD medium. Systematic errors are not shown.}
\label{fig:phenix}
\end{center}
\end{figure}

%%%%%%%%%%%%%%%%%%%%%%%%%%%%%%%%%%%%%%%%%%%%%%%%%%%%%%%%%%%%%%%%%%%%
\section{Conclusion}\label{se:summary}
%%%%%%%%%%%%%%%%%%%%%%%%%%%%%%%%%%%%%%%%%%%%%%%%%%%%%%%%%%%%%%%%%%%%

Let us summarize what has been carried out here.\\

The multiplicity of gluons $N(\omega)$ radiated by a hard quark in a QCD medium has first been calculated from the BDMPS gluon spectrum in the soft limit ($\omega_c \ll E$). The analytic IR behavior of $N(\omega)$ for outgoing quarks is recalled while we give similar expressions considering incoming quarks as well. These are shown to reproduce fairly well the exact multiplicity for gluon energies as large as the typical scale $\omega \sim \omega_c$.

Subsequently, this allows for the computation of the probability distribution from the integral representation given in Ref.~\cite{Baier:2001yt}. The quenching weight determined for both incoming and outgoing quarks is then compared to analytic estimates based on the small energy behavior of the gluon multiplicities. In particular, we have emphasized that the analytic expression given by BDMS strongly overestimates the high energy tail of the energy loss distribution, while an analytic (although complicated) expression based on the multiplicity $N(\omega \le \omega_c)$ (Eq.~(\ref{eq:Nsmall_out})) reproduces fairly well the full result. Noticing that the quenching weight follows a log-normal distribution, we give $D(\eps)$ a simple analytic parameterization.

Going a step beyond in Section~\ref{se:approx}, the probability distribution is computed from the BDMPS spectrum truncated in the IR ($\omega > \omega_{min}$) to ensure the LPM regime to be at work. It exhibits a ``discrete'' behavior at small energies (a few times the gluon energy cutoff) which corresponds to the emission of a small number of gluons, followed by a smoother ``continuum'' similar to what was found earlier. 

The $\cO{\omega/E}$ corrections in the gluon spectrum are then explicitly taken into account. This leads to a reduction of hard gluon emission which suppresses the tail of the probability distribution. To be complete, we give a (log-normal) parameterization for the quenching weight $D(\eps,E)$ for finite quark energy, which can be of use for future tomographic studies.

In conclusion, we illustrate the former results computing the quenching of hadron spectra in nuclear matter as well as in an expanding quark-gluon plasma. These estimates are respectively compared to HERMES $e$-A data and recent $\pi^0$ measurements by the PHENIX collaboration in Au-Au collisions at RHIC. While the HERMES data can quantitatively be understood as coming from the effect of quark energy loss, the trend observed in the PHENIX data is opposite to what one could naively expect. Finally, we suggest that this could be due to the fact that the ``hard'' quark produced at moderate $p_\perp$ actually hadronizes inside the medium. This has of course to be further investigated and is under current study.

\acknowledgments

I would like to express my gratitude to A.~Accardi, R.~Baier, Yu.~Dokshitzer, E.~Kolomeitsev, S.~Peign\'e, C.~Salgado, D.~Schiff, D.~Sousa, and U.~Wiedemann for many interesting and stimulating discussions about this work. It is also my pleasure to thank V.~Muccifora (HERMES) and D. d'Enterria (PHENIX) for discussions about the experimental data.

\providecommand{\href}[2]{#2}\begingroup\raggedright\endgroup

\end{document}